\documentclass[fleqn,twoside]{article}
\usepackage{amsfonts}

\def\noi{\noindent}

\makeatletter

\renewcommand{\thesubsubsection}%
        {\arabic{section}.\arabic{subsection}.\arabic{subsubsection}.}
\renewcommand{\paragraph}{\@startsection{paragraph}{4}{0pt}%
        {1.2ex plus -.5ex minus -.2ex}{-1em}{\normalsize\bf}}
\renewcommand{\@oddhead}{\raisebox{0pt}[\headheight][0pt]{%
   \vbox{\hbox to\textwidth{\rightmark \hfil \rm \thepage \strut}\hrule}}}
\renewcommand{\@evenhead}{\raisebox{0pt}[\headheight][0pt]{%
   \vbox{\hbox to\textwidth{\thepage \hfil \leftmark \strut}\hrule}}}
\newcommand{\heads}[2]{\markboth{\protect\small\it #1}{\protect\small\it #2}}

\makeatother

\newcommand{\Title}[1]{\noi {\Large #1} \\}
\newcommand{\Author}[2]{\noi{\large\bf #1}\\[2ex]\noi{\it #2}\\}
\newcommand{\Abstract}[1]{\vskip 2mm \begin{center}
     \parbox{16.4cm}{\small\noi #1} \end{center}\bigskip}
\newcommand{\foom}[1]{\protect\footnotemark[#1]}

\newcommand{\email}[2]{\footnotetext[#1]{e-mail: #2}
	\addtocounter{footnote}{1}}
\def\sect{Sec.\,}
\def\para{\paragraph}

\def\nqq{\hspace{-2em}}
\def\nhq{\hspace{-0.5em}}

\def\cm{\hspace{1cm}}
\def\inch{\hspace{1in}}

\def\eq{Eq.\,}
\def\eqs{Eqs.\,}
\def\beq{\begin{equation}}
\def\eeq{\end{equation}}
\def\bear{\begin{eqnarray}}
\def\al{&\nhq}
\def\lal{&&\nqq {}}               % left alignment
\def\bearr{\begin{eqnarray} \lal}
\def\ear{\end{eqnarray}}
\def\earn{\nonumber \end{eqnarray}}
\def\tst{\textstyle}
\def\dst{\displaystyle}

\def\nn{\nonumber\\ {}}

\def\nnn{\nonumber\\ \lal }
\def\nnnv{\nonumber\\[5pt] \lal }
\def\yy{\\[5pt]}

\def\eql{\al =\al}

\def\eqdef{\stackrel{\rm def}{=}}
\def\e{{\,\rm e}}
\def\d{\partial}

\def\sign{\mathop{\rm sign}\nolimits}
\def\diag{\mathop{\rm diag}\nolimits}

\def\const{{\rm const}}
\def\Half{{\dst\frac{1}{2}}}
\def\half{{\tst\frac{1}{2}}}
\def\then{\quad \Rightarrow \quad}

\def\ten#1{\mbox{$\cdot 10^{#1}$}}
\newcommand{\vars}[1]{\left\{\begin{array}{ll}#1\end{array}\right.}

\newcommand{\Theorem}[2]{\medskip\noi {\bf #1. \ }{\sl #2}\medskip}

\def\eps{\varepsilon}
\def\ep{\epsilon}
\def\o{\omega}

\def\ep{\epsilon}
\def\o{\omega}
\def\M{{\mathbb M}\,{}}
\def\Mph{\M_{\rm phys}}

\def\R{{\mathbb R}}
\def\S{{\mathbb S}}
\def\cA{{\cal A}}

\def\cS{{\cal S}}
\def\HT{{\cal H}_{\rm T}}
\def\V{{\mathbb V}}
\def\oI{{\overline I}}
\def\oJ{{\overline J}}
\def\od{{\overline d}}

\def\uc{{\underline c}}
\def\vY{\vec Y{}}

\def\olam{\overline{\lambda}}
\def\ophi{\overline{\varphi}}
\def\hq{\widehat q{}}
\def\Som{\cS_{\o}}
\def\sumo{\sum_{\o}}
\def\summo{\sum_{\mu\in\Som}}
\def\sumn{\sum_{i=1}^{n}}
\def\sums{\sum_s}
\def\sumt{\sum_{i=2}^{n}}
\def\umx{u_{\max}}
\def\m{{\rm m}}

\def\Fei{F_{\e I}}
\def\Fmi{F_{\m I}}

\def\Ysq{Y_s^2}
\def\Yoq{Y_\o^2}
\def\Fei{F_{\e I}}
\def\Fmi{F_{\m I}}

\def\eqos{\quad \stackrel{\rm OS}{=}\quad }
\def\bopr{{\beta^1}'}
\def\boprr{{\beta^1}''}

\def\ds{ds^2_D}
\def\mn{_{\mu\nu}}

\def\MN{^{\mu\nu}}
\def\Ref#1{Ref.\,\cite{#1}}
\def\rank{\mathop{\rm rank}\nolimits}

\def\sph{spherically symmetric}
\def\ssph{static, spherically symmetric}
\def\bh{black hole}
\def\bhs{black holes}

\def\brane{$p$-brane}
\def\branes{$p$-branes}
\def\TH{T_{{}_{\rm H}}}

\heads
     {K.A. Bronnikov and V.N. Melnikov}
     {On observational predictions from multidimensional gravity}

\begin{document}
\thispagestyle{empty}
\rightline{\bf gr-qc/0103079}
\bigskip

\Title
     {\bf On observational predictions from multidimensional gravity}

\Author
     {K.A. Bronnikov\foom 1 and V.N. Melnikov\foom 2$^{(*)}$}
     {Centre for Gravitation and Fundam. Metrology, VNIIMS,
         	3-1 M. Ulyanovoy St., Moscow 117313, Russia;\\
     Institute of Gravitation and Cosmology, PFUR,
	        6 Miklukho-Maklaya St., Moscow 117198, Russia and\\
     (*) Depto. de Fisica, CINVESTAV, Mexico}

\Abstract
     {We discuss possible observational manifestations of \ssph\ solutions
     of a class of multidimensional theories of gravity, which includes
     the low energy limits of supergravities and superstring
     theories as special cases. We discuss the choice of a physical
     conformal frame to be used for the description of observations.
     General expressions are given for (i) the Eddington parameters
     $\beta$ and $\gamma$, characterizing the post-Newtonian gravitational
     field of a central body, (ii) \brane\ \bh\ temperatures in different
     conformal frames and (iii) the Coulomb law modified by extra
     dimensions. It is concluded, in particular, that $\beta$ and $\gamma$
     depend on the integration constants and can be therefore different for
     different central bodies. If, however, the Einstein frame is adopted
     for describing observations, we always obtain $\gamma=1$. The modified
     Coulomb law is shown to be independent of the choice of a 4-dimensional
     conformal frame. We also argue the possible existence of specific
     multidimensional objects, T-holes, potentially observable as bodies
     with mirror surfaces.  }

\email 1 {kb@rgs.mccme.ru}
\email 2 {melnikov@rgs.phys.msu.su}

\section{Introduction}

     The observed physical world is fairly well described by the
     conventional 4-dimensional picture. On the other hand, in theoretical
     physics, whose basic aims are to construct a ``theory of everything''
     and to explain why our universe looks as we see it and not otherwise,
     most of the recent advances are connected with models in dimensions
     greater than four: Kaluza-Klein type theories, 10-dimensional
     superstring theories, M-theory and their further generalizations. Even
     if such theories (or some of them) successfully explain the whole
     wealth of particle and astrophysical phenomenology, there remains a
     fundamental question of finding direct observational evidence of extra
     dimensions, which is of utmost importance for the whole human world
     outlook.

     Observational ``windows'' to extra dimensions are actively discussed
     for many years. Thus, well-known predictions from extra
     dimensions are variations of the fundamental physical constants on the
     cosmological time scale \cite{will}--\cite{mel6}. Such constants are,
     e.g., the effective gravitational constant $G$ and the fine structure
     constant $\alpha$. There exist certain observational data on
     $G$ stability on the level of $\Delta G/G \sim 10^{-11}\div 10^{-12}$
     y$^{-1}$ \cite{will,mel3,damr}, which restrict the range of viable
     cosmological models. Very recently some evidence was obtained from
     quasar absorption spectra, testifying the variability of $\alpha$:
     $\Delta \alpha/\alpha \sim -0.72\ten{-5}$ over the redshift range $0.5
     < z  < 3.5$ \cite{alpha-} (the minus sign means smaller $\alpha$ in the
     past).

     Some effects connected with waves in small compactified extra
     dimensions are also discussed \cite{zhuk}: it is argued that such
     excitations can behave as particles with a large variety of masses and
     contribute to dark matter or to cross-sections of usual particle
     interactions.

     Other possible manifestations of extra dimensions are monopole modes in
     gravitational waves, various predictions for standard cosmological
     tests and generation of the cosmological constant \cite{GIM}, and
     numerous effects connected with local field sources, some of them being
     the subject of the present paper. These include, in particular,
     deviations from the Newton and Coulomb laws \cite{M1, mel4, mel5,
     bm-ann} and properties of \bhs.

     It had been conventional, starting from the pioneering papers of
     Kaluza and Klein, to suppose that extra dimensions, if any, are not
     directly observable due to their tiny size and compactness.
     For about two years, however, an alternative picture, connected
     with the so-called ``brane world'' models, is being actively developed.
     This trend rests on the suggestion advanced in 1982--83
     \cite{akama,rubshap} that we may live in a domain wall, or
     brane, of 3 spatial dimensions, embedded in a higher-dimensional space,
     which is unobservable directly since our brane world is located in a
     potential ``trench'' and/or most of the types of matter are
     concentrated on this brane. The recent boom was apparently launched by
     the works of Randall and Sundrum \cite{ransum} who showed, in
     particular, a way of obtaining Newtonian gravity on the brane from a
     multidimensional model.  Since their publication hundreds of papers
     have appeared, with a diversity of specific models and predictions. We
     will not try here to review this vast trend since it seems too early to
     make conclusions:  while one is preparing a survey, tens of new works,
     drastically changing the picture, can appear. We would only mention an
     opportunity suggested by Maia and Silveira \cite{maia} well
     before the present outburst. These authors argued that near a black
     hole (BH) particles may gain sufficient gravitational energy to
     overcome the potential barrier confining them to four dimensions and
     can thus run away from our world.

     In this paper we discuss in some detail predictions from extra
     dimensions connected with local sources of gravity. Assuming that extra
     dimensions of variable size inside and around such sources (e.g.,
     stars, galaxies, \bhs) can affect various physical phenomena, one
     should apply multidimensional theories to describe these phenomena.

     In \sect 2 and 3 we present some exact \ssph\ solutions of a
     generalized field model \cite{IM0}--\cite{bim97}, associated with
     charged \branes\ and motivated by the bosonic sector of the low-energy
     field approximation of superstring theories, M-theory and their
     generalizations \cite{GSW}--\cite{khv}. Our model, however, is not
     restricted to known theories since it assumes arbitrary dimensions of
     factor spaces, arbitrary ranks of antisymmetric forms and an arbitrary
     number of scalar fields.

     Then, in \sect 4, we discuss the choice of the conformal frame (CF) in
     which observational predictions should be formulated. Since at present
     there is no generally accepted unified theory, in our approach we use a
     generalized model with arbitrary \branes\ in diverse dimensions and
     study the physical applications on the basis of exactly solvable models.
     Thus we do not fix the underlying fundamental theory and have
     no reason to prescribe a particular frame, therefore all further
     results are formulated in an arbitrary frame.

     In \sect 5 we derive the post-Newtonian (PN) approximation of the above
     solutions in order to designate possible traces of extra dimensions and
     \branes\ in the comparatively weak gravitational fields of the great
     majority of planetary and stellar systems, including binary pulsars.
     This section generalizes the results of previous papers \cite{im-cots,
     im-mank}. For \bhs, apart from the PN parameters which determine the
     motion of test bodies in their sufficiently far neighbourhood, there is
     one more potentially observable parameter, the Hawking temperature
     $\TH$, which is obviously important for small (e.g., primordial) \bhs\
     rather than those of stellar or galactic mass range. We discuss the
     expressions for $\TH$ for a variety of \bh\ solutions and their
     applicability in different conformal frames.

     \sect 6 and 7 describe such consequences of these (and many other)
     field models as the Coulomb law violation and the possible existence of
     new, purely multidimensional objects, $T$-holes \cite{br95-1, br95-2}.
     The results are briefly discussed in \sect 8.

\section{$D$-dimensional action and minisuperspace representation}

\subsection{The model}

     The starting point is, as in Refs.\,\cite{IM0}--\cite{bim97},
     \cite{bobs, br-jmp}, the model action for $D$-dimensional gravity with
     several scalar dilatonic fields $\varphi^a$ and antisymmetric
     $n_s$-forms $F_s$:
\beq                                                         \label{2.1}
     S = \frac{1}{2\kappa^{2}}
	             \int\limits_{\M} d^{D}z \sqrt{|g|} \biggl\{
     R[g]
	- \delta_{ab} g^{MN} \d_{M} \varphi^a \d_{N} \varphi^b
                                                    - \sum_{s\in \cS}
 	\frac{\eta_s}{n_s!} \e^{2 \lambda_{sa} \varphi^a} F_s^2
                  \biggr\},
\eeq
     in a (pseudo-)Riemannian manifold
     $\M = \R_u \times \M_{0} \times \ldots \times \M_{n}$ with the
     \ssph\ metric
\bear
     \ds  = g_{MN}dz^M dz^N \eql
               \e^{2\alpha^0} du^2 +                         \label{2.11}
    	             \sum_{i=0}^{n} \e^{2\beta^i} ds_i^2  \nn
   	  \eql
              \e^{2\alpha^0} du^2 + \e^{2\beta^0} d\Omega^2
	    - \e^{2\beta^1} dt^2 +
    	             \sum_{i=2}^{n} \e^{2\beta^i} ds_i^2.
\ear
     Here $u$ is a radial coordinate ranging in $\R_u \subseteq \R$;
     $ds_0^2 = d\Omega^2$ is the metric on a unit $d_0$-dimensional sphere
     $\M_0 = \S^{d_0}$; $t \in \M_1 \equiv \R_t$ is time;
     the metrics $g^i = ds_i^2$ of the ``extra" factor spaces
     ($i\geq 2$) are assumed to be $u$-independent, Ricci-flat and can have
     arbitrary signatures $\eps_i=\sign g^i$; $|g| = |\det g_{MN}|$ and
     similarly for subspaces; $F_s^2 = F_{s,\ M_1 \ldots M_{n_s}} F_s^{M_1
     \ldots M_{n_s}}$; $\lambda_{sa}$ are coupling constants; $\eta_s = \pm
     1$ (to be specified later); $s \in \cS$,  $a\in \cA$, where $\cS$ and
     $\cA$ are some finite sets.

     This formulation admits both spacelike and timelike extra dimensions.
     Models with multiple timelike dimensions were considered in a
     number of papers, e.g., \cite{im2t} and more recently in \cite{2times,
     khuri}.

     The ``scale factors" $\e^{\beta^i}$ and the scalars $\varphi^a$ are
     assumed to depend on $u$ only.

     The $F$-forms should also be compatible with spherical symmetry.
     A given $F$-form may have several essentially (non-permutatively)
     different components; such a situation is
     sometimes called ``composite $p$-branes"%
\footnote{There is an exception:  two components,
   having only one noncoinciding index, cannot coexist since in this case
   there emerge nonzero off-block-diagonal components
   of the energy-momentum tensor (EMT) $T_M^N$, while the
   Einstein tensor in the l.h.s. of the Einstein equations is
   block-diagonal.  See more details in \Ref {IM5}.}.
   For convenience, we will nevertheless treat essentially different
   components of the same $F$-form as individual (``elementary") $F$-forms.
   A reformulation to the composite ansatz, if needed, is straightforward.

   The $n_s$-forms $F = F_{[n_s]} = dA_s$, where $A_s$ is a potential
   $(n_s-1)$-form, are naturally classified as
   {\it electric\/} ($\Fei$) and {\it magnetic\/} ($\Fmi$) ones, both
   associated with a certain subset $I = \{i_1, \ldots, i_k \}$ ($i_1 <
   \ldots < i_k$) of the set of numbers labeling the factor spaces:  $\{i\}
   = I_0 = \{0, \ldots, n \}$. By definition, the potential $A_I$ of an
   electric form $\Fei$ carries the coordinate indices of the subspaces
   $\M_i,\ i\in I$ and is $u$-dependent (since only a radial component of
   the field may be nonzero), whereas a magnetic form $\Fmi$ is built as a
   form dual to a possible electric one associated with $I$.
%   , so that
%\[
%   \Fei = dA_I \equiv
%   	      \d_{[M_1} A_{M_2\ldots M_{n_s}]}dz^{M_1}\ldots dz^{M_{n_s}}
%   \cm\ \Fmi = * dA_I,
%\]
%   where the asterisk denotes the Hodge operator in $\M$.
   Thus nonzero components of $\Fmi$ carry coordinate indices of the
   subspaces $\M_i,\ i\in \oI \eqdef I_0 \setminus I$. One can write:
\beq
    n_{\e I} = \rank F_{\e I} = d(I) + 1,\cm
    n_{\m I} = \rank F_{\m I} = D - \rank F_{\e I} = d(\oI)  \label{2.22}
\eeq
    where $d(I) = \sum_{i\in I} d_i$ are the dimensions of the subspaces
    $\M_I = \M_{i_1} \times \ldots \times \M_{i_k}$.
    The index $s$ will be used to jointly describe the two types of forms,
    so that \cite{bim97, bobs}
\beq                                                     \label{2.23}
     \cS = \{s\} = \{\e I_s\} \cup \{\m I_s\}.
\eeq

    We will make some more natural assumptions:

\begin{description}\itemsep -.5ex
\item [(i)]
    The branes only ``live'' in extra dimensions, i.e.,
    	$0 \not\in I_s,\ \  \forall s$.
\item[(ii)]
    The energy density of each $F$-form is positive:
    	$-T^t_t (F_s) > 0,\ \ \forall s$.
\end{description}

    When all extra dimensions are spacelike, the second requirement holds if,
    as usual, in (\ref{2.1}) $\eta_s = 1$ for all $s$.  In more general
    models, with arbitrary $\eps_i$, the requirement $-T^t_t > 0$ holds if
\bearr
     \eta_{\e I} = - \eps(I)\eps_t(I),\cm
     \eta_{\m I} = - \eps(\oI) \eps_t(\oI),
     						\label{2.25}\\ \lal
     \eps(I)\eqdef \prod_{i\in I}\eps_i, \cm
	\eps_t(I)= \vars{ 1, &  \R_t \subset \M_I,\\
                      -1  & \mbox {otherwise} }         \label{epst}
\ear
    where $\R_t$ is the time axis. If $\eps_t(I) =1$, we are dealing
    with a true electric or magnetic form, directly generalizing the
    familiar Maxwell field; otherwise the $F$-form behaves as an effective
    scalar or pseudoscalar in the physical subspace. $F$-forms with
    $\eps_t(I) =-1$ will be called {\sl quasiscalar.\/}

    Several electric and/or magnetic forms (with maybe different coupling
    constants $\lambda_{sa}$) can be associated with the same $I$ and are
    then labelled by different values of $s$.  (We sometimes omit the index
    $s$ by $I$ when this cannot cause confusion.)

    The forms $F_s$ are associated with \branes\ as extended sources of the
    \sph\ field distributions, where the brane dimension is $p=d(I_s)-1$,
    while $d(I_s)$ is the brane worldvolume dimension.

    The following example illustrates the possible kinds of $F$-forms.

\para{Example 1:} $D=11$ supergravity, representing the low-energy limit of
     M-theory \cite{brane}. The action (\ref{2.1}) for the bosonic sector of
     this theory (truncated by omitting the Chern-Simons term) does not
     contain scalar fields ($\varphi^a = \lambda_{sa} =0$) and the only
     $F$-form is of rank 4, whose various nontrivial components $F_s$
     (elementary $F$-forms, to be called simply $F$-forms according to the
     above convention) are associated with electric 2-branes [for which
     $d(I_s)=3$] and magnetic 5-branes [such that $d(I_s)=6$, see
     (\ref{2.22})]. The action has the form
\beq                                                            \label{11su}
     S = \frac{1}{2\kappa^{2}}
	             \int\limits_{\M} d^{11}z \sqrt{|g|} \biggl[
          R[g] - \frac{1}{4!} F_{[4]}^2
                  \biggr].
\eeq

     Let us put $d_0=2$ and ascribe to the external space-time coordinates
     the indices $M = u,\ \theta,\ \phi,\ t$ ($\theta$ and $\phi$ are the
     spherical angles), and let $M=2,\ldots,8$ refer to the extra
     dimensions.  Furthermore, let the extra factor spaces $\M_i$,
     $i=2,\ldots,8$, be one-dimensional and coincide with the respective
     coordinate axes.  The number $i=1$ is ascribed to the time axis, $\M_1
     = \R_t$, as stated previously.  Then different kinds of forms can be
     exemplified as follows:

\begin{description}\itemsep -.5ex
\item{}
     $F_{ut23}$ is a true electric form, $I = \{123\}$;
					 $\oI = \{045678\}$.
\item{}
     $F_{\theta\phi 23}$ is a true magnetic form, $I = \{145678\}$;
					 $\oI = \{023\}$.
\item{}
     $F_{u234}$ is an electric quasiscalar form, $I=\{234\}$;
					 $\oI = \{015678\}$.
\item{}
     $F_{\theta\phi t 2}$ is a magnetic quasiscalar form, $I=\{345678\}$;
						 $\oI = \{012\}$.
\end{description}

\subsection{The target space $\V$}

     Under the above assumptions, the system is well described using the
     so-called $\sigma$ model representation \cite{IM5}), to be briefly
     outlined here as applied to static, \sph\ systems.

     Let us choose, as in \cite{br-73} and many later papers, the harmonic
     $u$ coordinate in $\M$ ($\nabla^M \nabla_M u = 0$), such that
\beq                                                         \label{3.1}
     \alpha (u)= \sum_{i=0}^{n} d_i \beta^i
	         \equiv d_0\beta^0 + \sigma_1(u), \cm
      			    \sigma_1(u) \eqdef \sum_{i=1}^{n} d_i\beta^i.
\eeq

     The Maxwell-like field equations for $F_s$ may be integrated in a
     general form.  Indeed, for an electric form $F_s$ ($s = \e I$) the
     field equations due to (\ref{2.1})
\beq                                                        \label{Max}
     \d_u \Bigl(
     F_s^{utM_3\ldots M_m}\sqrt{|g|}\e^{2\lambda_{sa}\varphi^a}\Bigr)=0,
\eeq
     where $m=d(I)+1$, are easily integrated to give
\bearr
     F_s^{utM_3\ldots M_m} = Q_s                             \label{el}
      	\e^{-\alpha^0-\sigma_0-2\lambda_{sa}\varphi^a}
                      \eps^{M_3...M_{d(I)}}/\sqrt{|g_I|} \nnn \inch
		\ \then \
      \frac{1}{m!} F_s^2
      	= \eps(I) Q_s^2 \e^{-2\sigma(\oI)-2\olam_{s}\ophi}.
\ear
     where $\olam_{s}\ophi = \lambda_{sa}\varphi^a$,
     $\eps^{...}$ and $\eps_{...}$ are Levi-Civita symbols,
     $|g_I| = \prod_{i\in I}|g^i|$, and $Q_s= \const$ are charges.
     In a similar way, for a magnetic $m$-form $F_s$ ($s= \m I$,
     $m=d(\oI_s)$), the field equations and the Bianchi identities
     $dF_s=0$ lead to
\beq                                                         \label{mag}
     F_{s, M_1... M_{d(\oI)}} = Q_s
		\eps_{M_1...M_{d(\oI)}} \sqrt{|g_\oI|}
	\quad \then \quad
      \frac{1}{m!} F_s^2
      	= \eps(\oI) Q_s^2 \e^{-2\sigma(\oI) + 2\olam_{s}\ophi}
\eeq
     We use the notations
\beq
     \sigma_i = \sum_{j=i}^{n} d_j\beta^j (u), \cm       \label{sigma}
     \sigma (I) = \sum_{i\in I} d_i\beta^i (u).
\eeq

     Consequently, in the r.h.s. of the Einstein equations due to
     (\ref{2.1}), $R_M^N -\half \delta_M^N R = T_M^N$,
     the energy-momentum tensor (EMT) $T_M^N$ takes the form
\bearr                                                      \label{EMT}
    \e^{2\alpha}T_M^N
	  = -\half \sums \ep_s Q_s^2
	      	\e^{2\sigma(I) -2\chi_s\olam_s \ophi}
	      			\diag\bigl(+1,\ [+1]_I,\ [-1]_\oI \bigr)
     \ + \ \half \bigl(\dot\varphi{}^a)^2
	                        \diag(+1,\ [-1]_{I_0}\bigr)        \nnn
\ear
     where the first place on the diagonal belongs to $u$ and the symbol
     $[f]_J$ means that the quantity $f$ is repeated along the diagonal for
     all indices referring to $\M_j,\ j\in J$;
     $\sigma(I) \eqdef \sum_{i\in I} d_i\beta^i$;
     the sign factors $\ep_s$ and $\chi_s$ are
\beq
	\ep_{\e I} = -\eta_{\e I} \eps(I), \cm
	\ep_{\m I} = \eta_{\m I} \eps(\oI);\cm      	\label{3.14}
	\chi_{\e I}= +1, \cm
	\chi_{\m I}= -1,
\eeq
     so that $\chi_s$ distinguishes electric and magnetic forms.

     The positive energy requirement (\ref{2.25}) that fixes the input signs
     $\eta_s$, can be written as follows using the notations (\ref{3.14}):
\beq
     \ep_s = \eps_t(I_s).                      		\label{4.1}
\eeq
     Thus $\ep_s=1$ for true electric and magnetic forms $F_s$ and
     $\ep_s=-1$ for quasiscalar forms.

     Due to (\ref{EMT}), the combination ${u\choose u}
     + {\theta\choose \theta}$ of the Einstein equations, where $\theta$
     is one of the angular coordinates on $\S^{d_0}$, has a Liouville form,
     $\ddot \alpha -\ddot \beta^0 = (d_0-1)^2 \e^{2\alpha - 2\beta^0}$
     (an overdot means $d/du$), and is integrated giving
\bear
     \e^{\beta^0 - \alpha^0} = (d_0-1) s(k, u),\cm               \label{3.8}
     s(k,u)  \eqdef \vars{      k^{-1} \sinh ku, \quad & k>0,\\
     			     	             u,        & k=0,\\
     			  	k^{-1} \sin ku,        & k<0. }
\ear
     where $k$ is an integration constant. Another integration constant is
     suppressed by properly choosing the origin of $u$. With (\ref{3.8}) the
     $D$-dimensional line element may be written in the form
\beq
     \ds= \frac{\e^{-2\sigma_1/\od}}{[\od s(k,u)]^{2/\od}}
     	  \Biggl\{ \frac{du^2}{[\od s(k,u)]^2} + d\Omega^2\Biggr\}
     	      - \e^{2\beta^1} dt^2
	      	    + \sumt \e^{2\beta^i}ds_i^2,              \label{3.10}
\eeq
     $\od \eqdef d_0-1$.
     The range of the $u$ coordinate is  $0 < u < \umx$ where $u=0$
     corresponds to spatial infinity while $\umx$ may be finite or infinite
     depending on the form of a particular solution.

     The remaining set of unknowns ${\beta^i(u),\ \varphi^a (u)}$
     ($i = 1, \ldots, n,\ a\in \cA$) can be treated
     as a real-valued vector function $x^A (u)$ (so that
     $\{A\} = \{1,\ldots,n\} \cup \cA$) in an $(n+|\cA|)$-dimensional vector
     space $\V$ (target space). The field equations for $x^A$
     can be derived from the Toda-like Lagrangian
\bearr                                                      \label{3.12}
     L = G_{AB}\dot x^A\dot x^B - V_Q (y)
     \equiv \sumn d_i(\dot\beta^i)^2 + \frac{\dot\sigma_1^2}{d_0-1}
	          + \delta_{ab}\dot\varphi^a \dot\varphi^b - V_Q (y),\nnn
      V_Q (y) = -\sums \ep_s Q_s^2 \e^{2y_s}
\ear
     with the ``energy" constraint
\beq                                                      \label{3.16}
	E = G_{AB}\dot x^A \dot x^B + V_Q (y)
	                                = \frac{d_0}{d_0-1}\, k^2 \sign k,
\eeq
     where the IC $k$ has appeared in (\ref{3.8}).
     The nondegenerate symmetric matrix
\beq                                                       \label{3.13}
       (G_{AB})=\pmatrix {
  	       d_id_j/\od + d_i \delta_{ij} &       0      \cr
	          0                        &  \delta_{ab} \cr }
\eeq
     specifies a positive-definite metric in $\V$;
     the functions $y_s(u)$ are defined as scalar products:
\beq                                                       \label{3.15}
     y_s = \sigma(I_s) - \chi_s \olam_s\ophi
	   \equiv Y_{s,A}  x^A,    \cm\
     (Y_{s,A}) = \Bigl(d_i\delta_{iI_s}, \ \  -\chi_s \lambda_{sa}\Bigr),
\eeq
     where $\delta_{iI} =1$ if $i\in I$ and $\delta_{iI}=0$ otherwise.
     The contravariant components and scalar products of the vectors $\vY_s$
     are found using the matrix $G^{AB}$ inverse to $G_{AB}$:
\bearr                                                      \label{3.18}
     (G^{AB}) = \pmatrix{
	\delta^{ij}/d_i - 1/(D-2) &      0      \cr
	0                         &\delta^{ab}  \cr }, \cm\cm
	(Y_s{}^A) =
 	\Bigl(\delta_{iI}-\frac{d(I)}{D-2},
 				\quad -\chi_s \lambda_{sa}\Bigr); \nnn
				\\
\lal  Y_{s,A}Y_{s'}{}^A \equiv \vY_s \vY_{s'}
	                  = d(I_s \cap I_{s'})                \label{3.20}
     			      - \frac{d(I_s)d(I_{s'})}{D-2}
			      + \chi_s\chi_{s'} \olam_s \olam_{s'}.
\ear
     The equations of motion in terms of $\vY_s$ read
\beq
	\ddot{x}{}^A = \sums q_s Y_s{}^A \e^{2y_s},
	\cm    q_s  \eqdef \ep_s Q_s^2.                       \label{eqm}
\eeq

\section{Some exact solutions. Black holes}

\subsection{Exact solutions: orthogonal systems (OS)}

     The integrability of the Toda-like system (\ref{3.12}) depends on the
     set of vectors $\vY_s$, each $\vY_s$ consisting of input parameters of
     the problem and representing one of the $F$-forms, $F_s$, with a
     nonzero charge $Q_s$, in other words, one of charged \branes.

     In many cases general or special solutions to \eqs (\ref{eqm}) are
     known. The simplest case of integrability takes place
     when $\vY_s$ are mutually orthogonal in $\V$ \cite{bim97}, that is,
\beq                                                        \label{3.21}
     \vY_s \vY_{s'} = \delta_{ss'} Y_s^2, \cm
	     Y_s^2 =
	d(I)\bigl[1- d(I)/(D-2) \bigr] + \olam^2_{s} >0
\eeq
     where $\olam^2_s = \sum_a\lambda^2_{sa}$.
     Then the functions $y_s(u)$ obey the decoupled Liouville equations
     $\ddot y_s = \ep_s Q^2_s Y^2_s \e^{2y_s}$, whence
\beq                                       	            \label{3.23}
     \e^{-2y_s(u)} = \vars{
     	      Q^2_s Y^2_s\, s^2(h_s,\ u+u_s),            & \ep_s = 1,\yy
     	      Q^2_s Y^2_s h_s^{-2} \cosh^2 [h_s(u+u_s)], & \ep_s = -1,
						    \quad  h_s > 0,}
\eeq
     where $h_s$ and $u_s$ are integration constants and the function
     $s(.,.)$ has been defined in (\ref{3.8}). For the sought-for functions
     $x^A(u)$ and the ``conserved energy'' $E$ we then obtain:
\bear                                                        \label{3.24}
     x^A(u) \eql \sums \frac{Y_s{}^A}{\Ysq} y_s(u) + c^A u + \uc^A,
 \\
     E \eql \sums \frac{h_s^2\sign h_s}{\Ysq} + \vec c\,{}^2 \label{3.29}
                     = \frac{d_0}{d_0-1} \ k^2 \sign k,
\ear
     where the vectors of integration constants $\vec c$ and $\vec\uc$ are
     orthogonal to all $\vY_s$: \ $c^A Y_{s,A} = \uc^A Y_{s,A} = 0$, or
\beq
     c^i d_i\delta_{iI_s} - c^a\lambda_{sa}=0, \inch
     \uc^id_i\delta_{iI_s} - \uc^a\lambda_{sa}=0.     \label{3.25}
\eeq

\subsection{Exact solutions: block-orthogonal systems (BOS)}

     The above OS solutions are general for input parameters
     ($D$, $d_i$, $\vY_s$) satisfying \eq (\ref{3.21}): there is an
     independent charge attached to each (elementary) $F$-form.
     One can, however, obtain less general solutions for more general sets of
     input parameters, under less restrictive conditions than (\ref{3.21}).
     Namely, assuming that some of the functions $y_s(u)$ (\ref{3.15})
     coincide, one obtains the so-called BOS solutions \cite{bobs}, where
     the number of independent charges coincides with the number of
     different functions $y_s(u)$.

     Indeed, suppose \cite{bobs} that the set $\cS$ splits into several
     non-intersecting non-empty subsets,
\beq                                                      \label{*1}
     \cS = \bigcup_{\o}\Som, \cm |\Som|=m(\o),
\eeq
     such that the vectors $\vY_{\mu(\o)}$ ($\mu(\o) \in \Som$)
     form mutually orthogonal subspaces $\V_{\o} \subseteq \V$:
\beq
     \vY_{\mu(\o)} \vY_{\nu(\o')} = 0, \cm \o \ne \o'.    \label{*2}
\eeq
     Then the corresponding result from \cite{bobs} can be formulated as
     follows:

\Theorem{BOS solution}
    {Let, for each fixed $\o$, all $\vY_\nu \in \V_\o$
     be linearly independent, and let there be a vector
     $\vY_\o = \summo a_{\mu}\vY_{\mu}$ with $a_\mu \ne 0$ such that
\beq                                                            \label{*3}
     \vY_\mu \vY_\o = Y_\o^2 \eqdef \vY_\o^2,\cm \forall\ \mu\in\Som.
\eeq
     Then one has the following solution to the equations of motion
     (\ref{eqm}), (\ref{3.16}):
\bear                                                         \label{*6}
     x^A \eql \sum_\o \frac{Y_\o{}^A}{\Yoq} y_\o(u)+ c^A u + \uc^A,\\
							      \label{*7}
     \e^{-2y_{\o}}
         \eql \vars{
	 	\hq_{\o}  \Yoq s^2 (h_\o, u+u_\o), & \hq_\o >0,\\
	       |\hq_{\o}| \Yoq h_\o^{-2}\cosh^2 [h_\o(u+u_\o)],
		                           & \hq_\o < 0,\quad h_\o>0;}
     \ \ \
		\hq_\o \eqdef \sumo \ep_s Q^2_\mu,  \\
							      \label{*5}
       E \eql \sum_\o \frac{h_\o^2\sign h_\o}{\Yoq} + \vec c\,{}^2
                     			= \frac{d_0}{d_0-1}\ k^2 \sign k,
\ear
     where $h_\o$, $u_\o$, $c^A$ and $\uc^A$ are integration constants;
     $c^A$ and $\uc^A$ are constrained by the orthogonality
     relations (\ref{3.25}) \ {\rm (the vectors $\vec c$
     and $\vec{\uc}$ are orthogonal to each individual vector $\vY_s\in\V$);
     the function $s(.,.)$ has been defined in (\ref{3.8}).}
     }

     \eqs (\ref{*3}) form a set of linear algebraic equations with respect
     to the ``charge factors'' $a_\nu = \ep_\nu Q_\nu^2/\hq_{\o} \ne 0$,
     satisfying the condition $\sumo a_{\mu}=1$. The existence of a solution
     to (\ref{*3}) guarantees that $\hq_\o \ne 0$. On the other hand,
     if a solution to (\ref{*3}) gives $a_\mu=0$ for some
     $\mu\in \Som$, this means that the block cannot contain such a
     \brane, and then the consideration may be repeated without it.%
\footnote
  {Geometrically, the vector
  $\vY_{\o}$ solving \eqs (\ref{*3}) is the altitude of the pyramid formed by
  the vectors $\vY_{\mu}$, $\mu \in \Som$ with a common origin. The
  condition $a_\mu >0$ means that this altitude is located inside the
  pyramid, while $a_\mu=0$ means that the altitude belongs to one of its
  faces.
  \label{foogeom} }

     The function $y_{\o}(u)$ is equal to $y_{\mu(\o)}(u)=
     Y_{\mu(\o),A}x^A$, which is, due to (\ref{*3}), the same for all
     $\mu\in\Som$. The BOS solution generalizes the OS one, (\ref{3.23}),
     (\ref{3.24}): the latter is restored when each block contains a single
     $F$-form.

     Both kinds of solutions are asymptotically flat, and it is
     natural to normalize the functions $y_s(u)$ and $y_\o(u)$ by the
     condition $y_s(0)=0$ or $y_\o(0)=0$, so that the constants $u_s$ and
     $u_\o$ are directly related to the charges.

     Other solutions to the equations of motion are known, connected with
     Toda chains and Lie algebras \cite{IMmapa, im9901, im9910}.

\subsection {Black-hole solutions}

     Black holes (BHs) are distinguished among other \sph\ solutions by the
     existence of horizons instead of singularities in the physical
     4-dimensional space-time $\Mph$; the extra dimensions and scalar
     fields are also required to be well-behaved on the horizon to provide
     regularity of the $D$-dimensional metric. Thus BHs are described by
     the above solutions under certain constraints upon the input
     and integration constants. The no-hair theorem of \Ref {br-jmp} states
     that BHs are incompatible with quasiscalar $F$-forms. This means that
     all $\ep_s=1$, hence, in particular, in the above BOS solution
     (\ref{*3})--(\ref{*5}), $\hq_\o >0$ and all $a_\nu >0$.
     Furthermore, requiring that all the scale factors $\e^{\beta^i}$ (except
     $\e^{\beta^1} = \sqrt{|g_{tt}|}$ which should tend to zero) and
     scalars $\varphi^a$ tend to finite limits as $u\to \umx$, we get
     \cite{bobs}:
\bear
     h_{\o} = k >0, \cm \forall\  \o; \inch
     c^A = k \sumo Y_{\o}^{-2} Y_{\o}{}^A - k \delta^A_1       \label{5.4}
\ear
     where $A=1$ corresponds to $i=1$ (time). The constraint (\ref{3.29})
     then holds automatically. The value $u = \umx = \infty$ corresponds to
     the horizon. The same condition for the OS solution
     (\ref{3.23})--(\ref{3.25}) is obtained by replacing $\o \mapsto s$.

     Under the asymptotic conditions $\varphi^a \to 0$, $\beta^i \to 0$
     as $u\to 0$, after the transformation
\bearr
     \e^{-2ku} = 1 - \frac{2k}{\od r^\od},\cm \od \eqdef d_0-1  \label{5.5}
\ear
     the metric (\ref{3.10}) for BHs and the corresponding scalar fields
     may be written as
\bearr
     \ds=
     \biggl(\prod_{\o}H_{\o}^{A_{\o}}\biggr)\Biggl[-dt^2
    	  \biggl(1-\frac{2k}{\od r^\od}\biggr)\prod_{\o} H_{\o}^{-2/Y_\o^2}
\nnn \inch +
       \biggl(\frac{dr^2}{1-2k/(\od r^\od)} + r^2 d\Omega^2\biggr)
          + \sum_{i=2}^{n} ds_i^2
                       \prod_\o H_{\o}^{A_{\o}^i}\Biggr]; \nnnv  \label{5.7}
\cm  A_{\o} \eqdef \frac{2}{\Yoq}
                  \summo \frac{a_\mu d(I_{\mu})}{D-2}
	                        \eqos \frac 2{\Ysq}\frac{d(I_s)}{D-2}; \nnn
\cm
     A_{\o}^i \eqdef -\frac{2}{\Yoq}
                  \summo a_\mu\delta_{iI_\mu}
			   \eqos -\frac{2}{\Ysq} \delta_{iI_s};   \yy  \lal
     \varphi^a = -\sum_{\o} \frac{1}{Y^2_{\o}} \ln H_{\o}
		    		\summo a_\mu \lambda_{\mu a}
              \eqos -\sums \frac{\lambda_{sa}}{\Ysq} \ln H_s,    \label{phi}
\ear
     where $\eqos$ means ``equal for OS, with $\o\mapsto s$",
     and $H_{\o}$  are harmonic functions in $\R_+ \times \S^{d_0}$:
\beq                                                            \label{5.8}
     H_{\o} (r) =  1 + P_{\o}/(\od r^\od), \cm
		   P_{\o} \eqdef \sqrt{k^2 + \hq_\o \Yoq} - k.
\eeq

     The subfamily (\ref{5.4}), (\ref{5.7})--(\ref{5.8}) exhausts all BOS BH
     solutions with $k>0$; the OS ones are obtained in the special case of
     each block $\Som$ consisting of a single element $s$.
     The only independent integration constants remaining in BH solutions
     $k$, related to the observed mass (see below), and the brane charges
     $Q_s$.

\para{Example 2.}
     The simplest, single-brane BH solutions are described by (\ref{5.7}),
     (\ref{5.8}) where all sums and products in $s$ consist of a single
     term. These solutions are well known \cite{horst91}.
     the metric (\ref{2.11}) for,
     e.g., $D=11$ supergravity (\ref{11su}) can be presented as
\beq
     ds_{11}^2 = H^{d(I)/9}                                    \label{ds11}
      \biggl[ -\frac{1-2k/(\od r^{\od})}{H}\,dt^2
              + \biggl(\frac{dr^2}{1-2k/(\od r^\od)} + r^2 d\Omega^2\biggr)
	      + H^{-1} ds_{\rm on}^2 + ds_{\rm off}^2
	      		\biggr]
\eeq
     where $H = H(r) = 1 + P/(\od r^\od)$, $P = \sqrt{k^2 + 2Q^2} - k$,
     $\od = d_0-1$; $ds_{\rm on}^2$ and $ds_{\rm off}^2$ are
     $r$-independent ``on-brane'' and ``off-brane'' extra-dimension line
     elements, respectively; the dimension $d_0$ of the sphere $\M_0$
     varies from 2 to 7 for $d(I)=3$ (an electric brane) and from 2 to 4
     for $d(I)=6$ (a magnetic brane). In particular, the cases of maximum
     $d_0$, when off-brane extra dimensions are absent, correspond in the
     extremal near-horizon limits to the famous structures $AdS_4 \times
     \S^7$ (electric) and $AdS_7 \times \S^4$ (magnetic).  All these BH
     solutions are stable under linear \sph\ perturbations \cite{bm-stab};
     though, small multidimensional BHs, whose horizon size is of the order
     of the conpactification length, are known to possess the
     Gregory-Laflamme instability \cite{glaf94} related to distortions in
     extra dimensions.

     The above relations describe non-extremal BHs.
     Extremal ones, corresponding to minimum BH mass for given charges
     (the so-called BPS limit), are obtained in the limit $k \to 0$. The
     same solutions follow directly from (\ref{*5})--(\ref{*7}) under
     the conditions $h_{\o} = k = c^A =0$. For $k = 0$, the solution is
     defined in the whole range $r>0$, while $r=0$ in many cases
     corresponds to a naked singularity rather than an event horizon, so
     that we no more deal with a \bh.  However, in many other important
     cases $r=0$ is an event horizon of extremal Reissner-Nordstr\"om type,
     with an AdS near-horizon geometry and even the global metric turns out
     to be regular, as it happens for the $AdS_4 \times \S^7$ and $AdS_7
     \times \S^4$ structures mentioned in the previous paragraph
     \cite{brane};

  %% Some examples are mentioned in the Appendix.

     Other families of solutions, mentioned at the end of the previous
     section, also contain BH subfamilies. The most general BH solutions
     are considered in \Ref {im9910}.

\section{4-dimensional conformal frames}

     To discuss possible observational manifestations of the above
     solutions, it is necessary to specify the 4-dimensional physical
     metric. (Here and henceforth we put $d_0 = 2$.) A straightforward
     choice of the $\Mph$ section $g\mn$ of (\ref{2.11}) is not properly
     justified: there remains a freedom of multiplying this 4-metric by a
     conformal factor depending on the dilatonic fields $\varphi^a$ and the
     scale factors $\e^{\beta^i}$, $i \geq 2$. This is the well-known
     problem of the choice of a physical conformal frame (CF). Although
     mathematically a transition from one CF to another is only a
     substitution in the field equations, which can be solved using any
     variables, physical predictions, concerning the behaviour of massive
     matter, are CF-dependent.

     In \eq (\ref{2.1}), as well as in the previous sections, the
     $D$-dimensional Einstein (D-E) frame was used, although in
     such a general setting of the problem there is no evident reason to
     prefer one frame or another. The Einstein frame in various dimensions
     is distinguished by its convenience for solving the field
     equations due to the constant effective gravitational coupling.
     Due to the presence of $\sqrt{|g|}$ in the action, Einstein-frame
     metrics in different dimensions differ by certain volume factors. In
     particular, the 4-E metric for the theory (\ref{2.1}) is
\beq
     g^{\rm E}\mn  = \e^{\sigma_2}g\mn                     \label{4-E}
\eeq
     where $g\mn$ is the 4-dimensional part of the original metric
     $g_{MN}$ used in (\ref{2.1}), while $\e^{\sigma_2}$, defined in
     (\ref{sigma}), is the volume factor of all extra dimensions.

     The choice of a physical CF in non-Einsteinian theories of gravity is
     widely discussed, but the discussion is mostly restricted to the
     4-dimensional metric --- see e.g. \cite{sokol93,fara98,rzhuk98} and
     numerous references therein. There are arguments in favour of the
     Einstein frame, and the most important ones, applicable to higher order
     and scalar-tensor theories (and many multiscalar-tensor theories
     obtainable from multidimensional gravity) are connected with the
     positivity of scalar field energy and the existence of a classically
     stable ground state \cite{sokol93,fara98}; though, these requirements
     are violated if quantum effects are taken into account \cite{mel4}.

     In our view, however, the above arguments could be convincing if
     we were dealing with an ``absolute'', or ``ultimate'' theory of
     gravity. If, on the contrary, the gravitational action is obtained as a
     certain limit of a more fundamental unification theory, theoretical
     requirements like the existence of a stable ground state should be
     addressed to this underlying theory rather than its visible
     manifestation. In the latter, the notion of a physical CF should be
     only related to the properties of instruments used for measuring
     lengths and time intervals. Moreover, different sets of instruments
     (different measurement systems \cite{mel3}) are described, in general,
     by different CFs.

     Therefore, for any specific underlying theory that leads to the action
     (\ref{2.1}) in a weak field limit, two CFs are physically
     distinguished:  one, which may be called the {\it fundamental frame\/},
     where the theory is originally formulated and another one, the {\it
     observational frame,\/} or the {\it atomic system of measurements}
     (the 4-A frame), providing the validity of the weak equivalence
     principle (or geodesic motion) for ordinary matter in 4 dimensions.
     The fundamental frame is specified in the
     original space-time where the theory is formulated and is a natural
     framework for discussing such issues as space-time singularities,
     horizons, topology, etc.  ({\it what happens as a matter of fact\/}).
     On the other hand, the 4-A frame is necessary for formulating
     observational predictions ({\it what we see\/}), and its choice depends
     on how fermions are introduced in the underlying theory \cite{mel3,
     br95-2}. The reason is that as long as clocks and other instruments
     used in observations and measurements consist of fermionic matter, the
     basic atomic constants are invariable in space and time by definition.
     For instance, the modern definition of reference length is connected
     with a certain spectral line, determined essentially by the Rydberg
     constant and, basically, by the electron and nucleon masses.

     The (4-dimensional section of the) fundamental frame and the 4-A
     frame are, generally speaking, different, and none of them necessarily
     coincides with the 4-E frame, which represents the {\it gravitational
     system of measurements\/} \cite{mel3}.

     If the underlying theory is string theory, the fundamental frame is
     realized by the so-called ``string metric'' (see e.g.,
     \cite{banks, shira}), connected with $g_{MN}$ of \eq (\ref{2.1}) (the
     D-E metric) by a dilaton-dependent conformal factor. On the other hand,
     to distinguish the observational frame, we have to take into account
     that even for a fixed underlying theory, such frames may be different
     for different particular cosmological models. Thus, for the case of
     string theory, new results on the equivalence of quantum and some
     classical dilatonic brane-worlds in string and Einstein frames have
     been obtained in \Ref{noji01}. For a more general context of string
     theory, let us recall that, in the effective field-theoretic limit of
     string theory in 10 dimensions, the Lagrangian is presented in a form
     similar to (\ref{2.1})  with some quadratic fermion terms do not
     contain the dilaton field (\cite{GSW}, \eq (13.1.49)). If those terms
     are associated with matter, then, by analogy, it is reasonable for
     illustration purposes to write the matter Lagrangian $L_m$ in our
     generalized model simply as an additional term in the brackets of \eq
     (\ref{2.1}).

     In the observational frame, the matter part of the 4-dimensional action
     in terms of the corresponding metric $g^*_{\mu\nu}$ should be simply
     $\int d^4 x\sqrt{g^*} L_m$. Then $g^*\mn$ is related to $g\mn$
     in the following way \cite{br95-2}:
\beq
      g^*\mn  = \e^{\sigma_2/2}g\mn.                        \label{g*}
\eeq

     In what follows, since we do not fix a particular underlying theory, we
     leave the 4-dimensional CF arbitrary and only single out
     some results corresponding to the choices (\ref{4-E}) and (\ref{g*}).

\section {Post-Newtonian parameters. Black-hole observables}

     One can imagine that some real astrophysical objects (stars,
     galaxies, quasars, black holes) may be described (perhaps
     approximately) by some solutions of multidimensional theory of gravity,
     i.e., are essentially multidimensional objects, whose structure is
     affected by charged \branes. (It is in this case unnecessary to assume
     that the antisymmetric form fields are directly observable, though one
     of them may manifest itself as the electromagnetic field.)

     The post-Newtonian (PN) (weak gravity, slow motion) approximation of
     these multidimensional solutions then determines the predictions of the
     classical gravitational effects: gravitational redshift, light
     deflection, perihelion advance and time delay (see \cite{will,damr}).
     Observational restrictions on the PN parameters will then determine the
     admissible limits of theoretical models.

     For \sph\ configurations, the PN metric is conventionally written in
     terms of the Eddington parameters $\beta$ and $\gamma$ in isotropic
     coordinates, in which the spatial part is conformally flat \cite{will}:
\beq
      ds_{\rm PN}^2 =                                           \label{gPN}
      		- (1 -2V + 2\beta V^2) dt^2
		     + (1+ 2 \gamma V) (d\rho^2 + \rho^2 d\Omega^2)
\eeq
     where $d\Omega^2$ is the metric on $\S^2$,  $V = GM/\rho$ is the
     Newtonian potential, $G$ is the Newtonian gravitational constant and
     $M$ is the active gravitating mass.

     Observations in the Solar system lead to
     tight constraints on the Eddington parameters \cite{damr}:
\bear							      \label{obs-be}
	\gamma \eql 0.99984 \pm 0.0003, \\
							      \label{obs-ga}
	\beta \eql 0.9998 \pm 0.0006.
\ear
     The first restriction is a result of over 2 million VLBI observations
     \cite{Reas}. The second one follows from the $\gamma$ data and an
     analysis of lunar laser ranging data. In this case a high precision
     test based on the calculation of the combination $(4\beta - \gamma -
     3)$, appearing in the Nordtvedt effect \cite{Nor}, is used \cite{Dik}.

     For the multidimensional theory under consideration, the metric
     (\ref{gPN}) should be identified with the asymptotics of the
     4-dimensional metric from a solution in the observational (4-A) frame.
     Preserving its choice yet undetermined, we can write according to
     (\ref{3.10}) with $d_0=2$:
\beq
     ds^*_4= \e^{2f(u)}                                          \label{g4}
         \biggl\{-\e^{2\beta^1} dt^2
                + \frac{\e^{-2\sigma_1}}{s^2(k,u)} \biggl[
			\frac{du^2}{s^2(k,u)} + d\Omega^2\biggr]\biggr\}
\eeq
     where $f(u)$ is an arbitrary function of $u$, normalized for
     convenience to $f(0) =0$. Recall that by our notations
     $\sigma_1 = \beta^1 + \sigma_2$, the function $s(k,u)$ is defined in
     \eq (\ref{3.8}), and spatial infinity takes place at $u=0$. The choice
     of the frame (\ref{g*}) means $f=\sigma_2/4$.
     The 4-E frame (\ref{4-E}) corresponds to $f=\sigma_2/2$.

     Passing to isotropic coordinates in (\ref{g4}) with the relations
\beq
     \frac{d\rho}{\rho} = -\frac{du}{s(k,u)},
					    \cm                \label{u-rho}
     \frac{du^2}{s^2(k,u)} + d\Omega^2 =
     			\frac{1}{\rho^2} (d\rho^2 + \rho^2 d\Omega^2),
\eeq
     one finds that for small $u$ (large $\rho$)
\[
     \frac{1}{\rho} = u\biggl[ 1 - \frac{u^2}{4}k^2 \sign k + O(u^4)\biggr],
\]
     so that $u=1/\rho$ up to cubic terms, and the decomposition in powers of
     $1/\rho$ up to $O(\rho^{-2})$, needed for comparison with (\ref{gPN}),
     precisely coincides with the $u$-decomposition near $u=0$.

     Using this circumstance, it is easy to obtain for the mass
     and the Eddington parameters corresponding to (\ref{g4}):
\beq
     GM = -\bopr - f';                                        \label{Ed-f}
     	\cm
     \beta = 1 + \Half\,\frac{\boprr + f''}{(GM)^2},
	\cm
     \gamma = 1 + \frac{2f' - \sigma_2'}{GM},
\eeq
     where $f' = df/du\Big|_{u=0}$ and similarly for other functions.
     The expressions (\ref{Ed-f}) are quite general, being applicable to
     asymptotically flat, \ssph\ solution of any theory where the EMT has
     the property $T^u_u + T^\theta_\theta =0$, which leads to the metric
     (\ref{3.10}).  They apply, in particular, to all solutions of the
     theory (\ref{2.1}) under the conditions specified, both mentioned and
     not mentioned above and those yet to be found.

     In the observational frame (\ref{g*}) we have
\beq
     GM = -\frac{1}{4} (3\bopr +\sigma'_2);               \label{Ed-f*}
     	\cm
     \beta = 1 + \frac{3\boprr + \sigma''_2}{8(GM)^2},
	\cm
     \gamma = 1 - \frac{\sigma'_2}{2GM}.
\eeq
     Similar expressions for the 4-E frame (\ref{4-E}) are
\beq
     GM = -\frac{1}{2} (2\bopr +\sigma'_2);               \label{Ed-E}
     	\cm
     \beta = 1 + \frac{2\boprr + \sigma''_2}{4(GM)^2},
	\cm
     \gamma = 1.
\eeq
     We thus conclude that {\it the Eddington parameter $\gamma$ is the same
     as in general relativity for all \brane\ solutions in the general model
     (\ref{2.1}) in the 4-E frame\/} (under the assumptions of \sect 2).

     Expressions of $\beta$ and $\gamma$ for specific solutions
     can be obtained by substituting them to (\ref{Ed-f}) or (\ref{Ed-f*}).
     One may notice, however, that $\beta$ may be calculated {\it directly
     from the equations of motion (\ref{eqm}), without solving them\/}.
     This is true for any function $f$ of the form $f = \vec F \vec x$ where
     $\vec F \in \V$ is a constant vector (i.e., $f$ is a linear combination
     of $\beta^i$ and the scalar fields $\varphi^a$):
\beq
     \beta - 1 = \frac{1}{2(GM)^2} \sums                   \label{be-f}
     		 \ep_s Q_s^2 (Y_s^1 + \vec F \vY_s) \e^{2y_s(0)}.
\eeq
     In particular, if $f = N\sigma_2$,
\beq
     \beta - 1 = \frac{1}{2(GM)^2}                            \label{be*}
		\Biggl\{
	\sum_{s: \ep_s{=}+1} Q_s^2 \biggl[
			1-N + \frac{(2N-1) d(I_s)}{D-2}\biggr]\e^{2y_s(0)}+
	\sum_{s: \ep_s{=}-1} Q_s^2 \frac{(1-2N)d(I_s)}{D-2}
						\e^{2y_s(0)}\Biggr\};
\eeq
     recall that $\ep_s=1$ refers to true electric and magnetic forms,
     $\ep_s=-1$ to quasiscalar ones. For $N=1/2$ and $N=1/4$ we obtain the
     values of $\beta$ for the frames (\ref{4-E}) and (\ref{g*}),
     respectively.

     Explicit expressions for $M$ and $\gamma$ (in frames other than
     4-E) require the asymptotic form of the solutions.

     It is convenient, without loss of generality, to normalize the
     scale factors at spatial infinity in such a way that $\e^{\beta_i (0)}
     = 1$, $i= 1, \ldots, n$, so that the real scales of the extra
     dimensions are hidden in the factor space metrics $g^i$. In a similar
     way, one can re-define the dilatonic fields: $\varphi^a - \varphi^a(0)
     \mapsto \varphi^a$, so that $\varphi^a(0)=0$, while the former
     asymptotic values of $\varphi^a$ have been actually absorbed in the
     charges $Q_s$. Then all $y_s(0)=0$.

     For all OS and BOS solutions it then follows that the constants $\uc^i$
     are zero; the constants $u_\o$, $h_\o$ and $\hq_\o$ in (\ref{*7})
     are related by
\beq
     1= \vars{ 	\hq_{\o}  \Yoq s^2 (h_\o, u_\o), & \hq_\o >0,\yy \label{asy1}
	       |\hq_{\o}| \Yoq h_\o^{-2}\cosh^2 (h_\o u_\o),
		                           & \hq_\o < 0,\quad h_\o>0.}
\eeq
     In a similar way for OS, according to (\ref{3.23}),
\beq                                       	               \label{asy2}
     1 = \vars{Q^2_s Y^2_s\, s^2(h_s,\ u_s),            & \ep_s = 1,\yy
     	       Q^2_s Y^2_s h_s^{-2} \cosh^2 (h_s u_s),  & \ep_s = -1,
						    \quad  h_s > 0.}
\eeq
     In (\ref{asy2}), the second line corresponds to a quasiscalar form
     $F_s$, while in (\ref{asy1}) the second line means that the summed
     squared charge $\hq_\o$ of the block $\Som$ is dominated by quasiscalar
     forms.

     In what follows we will only give expressions for BOS solutions; their
     OS versions are then evident.

     \eq (\ref{asy1}) gives
\beq                                                            \label{y'0}
     y'_\o = \frac{dy_\o}{du}\biggr|_{u=0} =
	  \vars{-
	        (\hq_\o \Yoq + h^2_\o \sign h_\o)^{1/2}, & \hq_\o >0;\yy
		\pm (h^2_\o - |\hq_\o| \Yoq)^{1/2},      & \hq_\o <0,    }
\eeq
     For OS $\hq_\o$ is replaced by $\ep_s Q^2_s$,
     the two lines refering to $\ep_s=1$ and $\ep_s=-1$, respectively.

     The quantities needed for calculating $M$ and $\gamma$ are
\bear
     x^A{}' = \sumo Y_\o^{-2}Y_\o^A y'_\o + c^A,         \label{bopr}
\cm
     \sigma'_1 = \Half \sumo A_\o y'_\o + \sum_{i=1}^{n}d_i c^i,
\ear
     with $A_\o$ defined in (\ref{5.7}). The quantity $\sigma_2$ can be
     obtained as $\sigma_1 - \beta^1$; as before, $x^1 = \beta^1$,
     $Y^1_\o = \summo a_\mu Y_\mu^1$ and
     $Y^1_\mu = \delta_{1I_\mu} - d(I_\mu)/(D-2)$.
     The values of $M$ and $\gamma$ are now easily found from
     (\ref{Ed-f}) in terms of the solution parameters for any given $f$ of
     the above general form, $f = \vec F \vec x$.

     In particular, for BH solutions (\ref{5.4})--(\ref{5.8}) there is no
     need to change the coordinates from $u$ to $r$ or $\rho$: it is
     sufficient to use \eqs (\ref{5.4}) for the constants. Moreover,
     BH solutions contain only true electric and magnetic forms, $\ep_s=+1$
     \cite{br-jmp} and $\hq_\o >0$.

     Thus, for instance, for BOS BH solutions the quantities $\bopr$ and
     $\sigma'_2$ are
\beq
     \bopr = -k - \sumo P_\o \frac{1-b_\o}{\Yoq}, \cm
     \sigma'_2 = -\sumo \frac{1 - 2b_\o}{\Yoq}                 \label{parBH}
\eeq
     with $P_\o$ defined in (\ref{5.8}) and $b_\o \eqdef \summo a_\mu
     d(I_\mu)/(D-2)$.  In the OS case $b_\o$ becomes $b_s=d(I_s)/(D-2)$.

     Accordingly, for BHs in CFs with $f= N\sigma_2$ we obtain
\bear
     GM = k + \sumo \frac{P_\o}{\Yoq}[1-b_\o + N(1-2b_\o)], \cm
     \gamma = \frac{1-2N}{GM}\sumo \frac{P_\o}{\Yoq}(1-2b_\o).\label{Mg-BH}
\ear

     Some general observations can be made from the above relations.
\begin{itemize}
\item
     The expressions for $\beta$ depend on the input constants $D$, $d(I_s)$
     (hence on \brane\ dimensions: $p_s = d(I_s) - 1$), on the mass $M$ and
     on the charges $Q_s$.  For given $M$, they are independent of other
     integration constants, emerging in the solution of the Toda system
     (\ref{eqm}), and also on \brane\ intersection dimensions, since they
     are obtained directly from \eqs (\ref{eqm}) \cite{im9901}. This means,
     in particular, that $\beta$ is the same for BH and non-BH
     configurations with the same set of input parameters, mass and charges.
\item
     According to (\ref{be*}), all \branes\ give positive contributions to
     $\beta$ in both frames (\ref{4-E}) and (\ref{g*}), therefore
     (\ref{be*}) combined with (\ref{obs-be}) leads to a general restriction
     on the charges $Q_s$ for given mass and input parameters.
\item
     The expressions for $\gamma$ depend, in general, on the integration
     constants $h_s$ or $h_\o$ and $c^i$ emerging from solving \eqs
     (\ref{eqm}). For BH solution these constants are expressed in terms of
     $k$ and the input parameters, so both $\beta$ and $\gamma$ depend on
     the mass, charges and input parameters.
\item
     In the 4-E frame, one always has $\gamma=1$. The same is true for some
     BH solutions in all frames with $f=N \sigma_2$. Indeed,
     a pair of electric and magnetic \branes\ with equal $|Q_s|$,
     corresponding to $F$-forms $F_1$ and $F_2$ of equal rank (in
     particular, if $F_1$ and $F_2$ are the electric and magnetic components
     of the same composite $F$-form), always forms a BOS block, with
     $a_1=a_2=1/2$, so that $b_\o=1/2$, and this pair does not contribute to
     $\sigma'_2$ in (\ref{parBH}). Evidently $\gamma=1$ as well for a BOS
     \bh\ containing several such dyonic pairs and no other $F$-forms.
     This property was noticed in \Ref{im-mank} for the frame $f=0$.
\end{itemize}

\noi{\bf BH temperature.}
     BHs are, like nothing else, strong-field gravitational objects, while
     the PN parameters only describe their far neighbourhood. An important
     characteristic of their strong-field behaviour, potentially observable
     and depending on their multidimensional structure, is the Hawking
     temperature $\TH$. As with other observables, it is of importance to
     know the role of conformal frames for its calculation. One can
     ascertain, however, that this quantity is {\it CF-independent\/}, at
     least if conformal factors that connect different frames are regular on
     the horizon.

     Indeed, if, in an arbitrary \ssph\ space-time with the metric
\beq
     g = -\e^{2\Gamma(r)}dt^2 + \e^{2{\rm A}(r)}dr^2     \label{sphm}
		+\mbox{anything else},
\eeq
     the sphere $r = r_{\rm hor}$ is an event horizon, its Hawking
     temperature can be calculated as
\beq                                                         \label{TH}
      \TH = \frac{1}{2\pi k_{\rm B}}
      	\lim_{r\to r_{\rm hor}} \e^{\Gamma-{\rm A}} |d\Gamma/dr|
\eeq
     where $k_{\rm B}$ is the Bolzmann constant.
     This expression \cite{bim97}, which is invariant under
     reparametrizations of the radial coordinate $r$, is easily obtained
     from standard ones \cite{wald}. The factor $\e^{\Gamma-{\rm A}}$
     is insensitive to conformal transformations $g \mapsto \e^{2f(r)}g$,
     whereas $\Gamma$ is replaced by $\Gamma + f$. At a horizon, $\Gamma\to
     -\infty$, and, if $r_{\rm hor}$ is finite (such a coordinate always
     exists), $|d\Gamma/dr|\to \infty$.  Therefore $\TH$ calculated
     according to \eq (\ref{TH}) will be the same in all frames with
     different $f(r)$ provided $df/dr$ is finite at $r=r_{\rm hor}$.

     This is precisely the case with the BH metric (\ref{5.7}) and any
     $f$ formed as a linear combination of $\beta^i$ ($i>1$ and $\varphi^a$.
     Using the recipe (\ref{TH}), one obtains \cite{bobs}
\beq
     \TH = \frac{1}{8\pi k k_{\rm B}}
		\prod_\o \biggl(\frac{2k}{2k+P_\o}\biggr)^{1/\Yoq}
         \eqos
	      \frac{1}{8\pi k k_{\rm B}}                      \label{THBOS}
		\prod_s \biggl(\frac{2k}{2k+P_s}\biggr)^{1/\Ysq}.
\eeq

     The physical meaning of $\TH$ is related to quantum evaporation, a
     process to be considered in the fundamental frame, while the produced
     particles with a certain spectrum are usually assumed to be observed at
     flat infinity, where our CFs do not differ. This means that the $\TH$
     expression should be CF-independent. We have seen that it possesses
     this property ``by construction''. The conformal invariance of $\TH$
     was also discussed in another context in \Ref{JaK}.

     All this is true for $\TH$ in terms of the integration constant $k$ and
     the charges $Q_s$. However, the observed mass $M$ as a function of the
     same quantities is frame-dependent, see (\ref{Ed-f}). Therefore $\TH$
     as a function of $M$ and $Q_s$ is frame-dependent as well. Thus, for
     small charges, $Q_s^2 \ll k^2$, one easily finds from (\ref{Mg-BH})
     under the same assumption $f = N \sigma_2$:
\beq
     k = GM -\frac{1}{2GM}\sumo \hq_\o [1-b_\o+N (1-2b_\o)]    \label{kM}
\eeq
     up to higher order terms in $\hq_\o/(GM)^2$ [$b_\o$ was defined in
     (\ref{parBH})]. This expression should be substituted into
     (\ref{THBOS}). For larger charges the corresponding expressions are
     more involved.

     The temperature of extremal BHs ($k=0$) only depends on the charges and
     the input parameters $\Yoq$. Moreover, by (\ref{THBOS}), for some sets
     of $\Yoq$, $\TH$ can tend to infinity as $k \to 0$. This means that the
     horizon becomes a naked singularity in the extremal limit \cite{bobs}.

\section {Coulomb law violation}

     One of specific potentially observable effects of extra dimensions is
     Coulomb law violation.

     Consider the space-time $\M$ described in \sect 2, with the metric
     (\ref{2.11}) and $d_0=2$.  Suppose that the electrostatic field
     of a \sph\ source is described by a term
     $\propto F^2 \e^{2\olam\ophi}$ in the action (\ref{2.1})
     (where, as before, $\olam\ophi = \lambda_a\varphi^a$),
     corresponding to a true electric $m$-form $\Fei$ with a certain
     set $I \ni 1$, or
\[
     I = {1} \cup J, \cm J \subset \{2,\ldots, n\}.
\]
     As before, from the field equations for $\Fei$ we have \eq (\ref{el}),
     so that
\beq
      \frac{1}{m!} F^2
          	= \eps(I) Q^2 \e^{-2\sigma(\oI)-2\olam\ophi}.    \label{QD}
\eeq

     The observable electromagnetic field $F\mn$ in 4 dimensions is
     singled out from $\Fei$ as follows:
\beq
     F_{\mu\nu M_3\ldots M_{m}} = F\mn,    	              \label{F2D}
		\cm
     \frac{1}{m!} F^2 = \Half\eps(J) F\mn F\MN \e^{-2\sigma(J)},
\eeq
     where the indices $M_3, \ldots M_{n_s}$ belong to $J_s$. $F\mn F\MN$ is
     written here in terms of the $g\mn$, the 4-dimensional part of
     $g_{MN}$. If, as in \sect 5, the observable metric is assumed to be
     $h\mn = \e^{2f} g\mn$, then the squared observed radial electric field
     strength is
\beq                                                          \label{F4}
     E^2 [f] = -h^{tt}h^{uu}(F_{ut})^2
             = -\e^{-4f}g^{tt}g^{uu}(F_{ut})^2 =
     					-\Half \e^{-4f}F\mn F\MN
\eeq
     since $F_{ut}=-F_{tu}$ is the only nonzero component of $F\mn$.
     From (\ref{F4}) with (\ref{F2D}) and (\ref{QD}) we obtain
\beq
     E^2 [f] = Q^2 \e^{-4f -4\beta^0}\cdot                    \label{E2f}
                    \e^{-4\olam\ophi + 2\sigma(J) - 2\sigma(\oJ)}
\eeq
     where $\e^{2\beta^0} = g_{\theta\theta}$, the notations (\ref{sigma})
     are used and $\oJ = \{2,\ldots, n\}\setminus J$.

     One can notice that $\e^{2f+2\beta^0} = h_{\theta\theta}= r^2$ where
     $r$ is the observable radius of coordinate spheres $t=\const,\
     u=\const$. Therefore \eq (\ref{E2f}) may be rewritten as
\beq                                                             \label{Ef}
     E = E [f] = (|Q|/r^2)\e^{-2\olam\ophi + \sigma(J) - \sigma(\oJ)}.
\eeq

     This is the modified Coulomb law.
     The deviations from the conventional Coulomb law are evidently both due
     to extra dimensions (depending on the multidimensional structure of the
     $F$-form) and due to the interaction with the scalar fields. This
     relation (generalizing the one obtained in \Ref{bm-ann} in the
     framework of dilaton gravity) is valid for an arbitrary metric of the
     form (\ref{2.11}) ($d_0=2$) and does not depend on whether or not this
     $F$-form takes part in the formation of the gravitational field.

     \eq (\ref{Ef}) is exact and --- which is remarkable --- it is {\it
     CF-independent}. This is an evident manifestation of the
     conformal invariance of the electromagnetic field in $\Mph$ even in the
     present generalized framework.

     From the observational viewpoint, the weak gravitational field
     approximation, in the spirit of the previous section, can be of
     interest. Consider for simplicity the conventional Maxwell field, i.e.,
     the 2-form $F_{MN}$ with the only nonzero component $F_{ut}=-F_{tu}$,
     so that $J = \emptyset$, $\sigma(J)=0$ and $\sigma(\oJ) = \sigma_2$.
     Under the assumptions of \sect 5, for large radii $r$ (small $u$)
\beq
     E  = (|Q|/r^2)\e^{-2\olam\ophi - \sigma_2}           \label{Eweak}
        = (|Q|/r^2)[1 -2\olam\ophi'/r - \sigma'_2/r + O(r^{-2})],
\eeq
     where expressions for $\ophi'$ and $\sigma'_2 = \sigma'_1 - \bopr$
     should be taken from specific solutions of the equations of motion
     --- see \eqs (\ref{y'0}), (\ref{bopr}).
     (Note that up to higher-order terms, $u\approx 1/\rho \approx 1/r$ at
     small $u$.) Since, in general, quantities like $\lambda\varphi'$ and
     $\sigma'_2$ are of the order of $k \sim GM$, one can conclude that the
     Coulomb law violation intensity is of the order of the gravitational
     field strength characterized by the ratio $GM/r$. Though,
     unlike $E$, the mass $M$ itself is $f$-dependent, see (\ref{Ed-f}).

\section{T-holes}

     Consider a simple example of a BH metric, e.g., (\ref{ds11}) in the
     case $d_0=2$, $Q=0$, hence $p=0$ and $H\equiv 1$. It is thus a direct
     generalization of the Schwarzschild metric, charged \branes\ are
     absent and all extra dimensions form a 7-dimensional manifold with the
     $r$-independent Ricci-flat metric $ds^2_7$, which we will assume to
     be flat. Let us introduce the following modification: interchange
     the time $t$ with a selected extra coordinate, say, $v$. One has
\beq
     ds_{11}^2 =                                           \label{ds-T}
              - dt^2
              + \frac{dr^2}{1-2k/r} + r^2 d\Omega^2
              + \biggl(1-\frac{2k}{r}\biggr)\eta_v \, dv^2
	      + ds_6^2
\eeq
     where $ds_6^2$ is the flat metric of the remaining dimensions and
     $\eta_v=\pm 1$ depending on whether the coordinate $v$ is spacelike or
     timelike.

     This modification only changes the interpretation of different
     coordinates without changing the mathematical properties of the metric,
     it therefore remains to be a solution of the field equations.

     The main feature of this configuration is that the physical space-time
     $\Mph$ changes its signature at $r = 2k$: it is $(+---)$ for $r > 2k$
     and $(++--)$ at $R < 2k$. This evidently means that the anomalous
     domains should be characterized with quite unconventional physics
     whose possible consequences and observational manifestations are yet
     to be studied. It has been suggested \cite{br95-1} to call such
     domains with an unusual space-time signature {\sl time holes} or {\sl
     T-holes} and the corresponding horizons {\sl T-horizons}, to be
     designated $\HT$.

     Evidently each BH configuration of any dimension $D > 4$
     has a family of T-hole counterparts (a family since the factor spaces
     may have different dimensions and signatures, and a $v$-axis like the
     one in (\ref{ds-T}) may be selected in any of them). Conversely, any
     T-hole solution has BH counterparts. If a BH possesses an external
     field, such as the $F_{[4]}$-form field corresponding to the metric
     (\ref{ds11}), under a BH---T-hole transition, its true
     electric or magnetic component may be converted into a quasiscalar one.
     This happens if the new $t$ coordinate (former $v$) is off-brane,
     $\R_t\not\in \M_I$. If the new time axis belongs to $\M_I$, the
     \brane\ remains true electric or magnetic.

     Unlike a BH-horizon, a T-horizon $\HT$ is not in absolute past or
     future from a distant observer's viewpoint, it is visible since it
     takes a finite time for a light signal to come from it (independently
     of a conformal gauge since the latter does not affect light
     propagation).

     Thus, in addition to the above family of \brane\ BH solutions,
     there is a similar family of T-hole ones.

     There are certain problems connected with the compactification
     of extra dimensions. They can be clearly understood using the
     simple example (\ref{ds-T}), which may be called the {\it
     T-Schwarzschild} metric. Note that if we ignore the ``passive''
     subspace with the metric $ds_6^2$, the remaining 5-dimensional
     manifold coincides with the ``zero dipole moment soliton'' in the
     terminology of \cite{groper}.

     At $r=2k$ the signs of $g_{rr}$ and $g_{vv}$ change simultaneously.
     Moreover, if $\eta_v =-1$, i.e., this compactified direction is
     timelike at large $r$, the total signature of $\M$ is preserved but
     in the opposite case, $\eta_v =-1$, it changes by four: two
     spacelike directions become timelike. However, as one can directly
     verify, $\HT$ is not a curvature singularity, either for the
     $D$-dimensional metric or for its 4-dimensional section.

     If $\eta_v=-1$, the surface $r=2k$ is a Schwarzschild-like horizon in
     the $(r,v)$ subspace, and there exists an analytic continuation to
     $R < 2k$ with the corresponding Kruskal picture. However, if some points
     on the $v$ axis are identified, as should be done to compactify the
     axis $\R_v$ in the conventional way, then the corresponding sectors
     are cut out in the Kruskal picture, so that the $T$-domain and
     $R$-domain sectors join each other only at a single point, the horizon
     intersection point. This should be probably interpreted as a
     singularity due to intersection of particle trajectories.

     Another thing happens if $\eta_v=1$. Again a further study is possible
     using a transition to coordinates in which the metric is manifestly
     nonsingular at $r=2k$. Let us perform it for (\ref{ds-T}) in the
     vicinity of $\HT$ (more general T-holes may be treated
     in a similar way):
\bearr
      r\to 2k; \qquad   r-2k=(x^2+y^2)/(8k);\qquad v=4k\arctan(y/x); \nnn
      ds^2_2 (r,v)
                  \approx \frac{r-2k}{2k}dv^2 + \frac{2k}{r-2k}dr^2
	= dx^2 + dy^2.   					\label{Tor}
\ear
     Thus the $(r,v)$ surface metric is locally flat near the T-horizon
     $r=2k$ which is transformed into the origin $x=y=0$, while the $v$
     coordinate has the character of an angle.

     This transformation could also be conducted as a conformal mapping
     of the complex plane with the aid of the analytic function $\ln z,\
     z=x+ {\rm i}y$, as was done in \Ref {br79} for some cylindrically
     symmetric Einstein-Maxwell solutions; then $v$ is proportional to \
     arg$\,z$.

     Consequently, in the general case the $(r,v)$ surface near $r=2k$
     behaves like a Riemann surface having a finite or infinite (if $v$
     varies in an infinite range) number of sheets, with a branch point
     at $x=y=0$ (a branch-point singularity \cite{br79}). If $\R_v$ is
     compactified, $v$ is naturally described as an angular coordinate
     ($0\leq v < 2\pi l$, where $v=0$ and $v=2\pi l$ are identified and
     $l$ is the compactification radius at the asymptotic $R\to\infty$).
     $r=2k$ is then the center of symmetry in the $(r,v)$ surface; the
     surface itself has the shape of a tube with a constant thickness at
     $r\to\infty$, becoming narrower at smaller $r$ and ending at $r=2k$
     either smoothly (if the regular center condition $l=4k$ is
     satisfied), or with a conical or branch-point singularity (otherwise).
     This suggests that there is no way to go beyond $r=2k$.

     In the singular case the geodesic completeness requirement is violated
     on $\HT$, so it is reasonable to require $l=4k$, or, more
     generally, $l=4kj$ where $j$ is a positive integer, so that $r=2k$
     is a $j$-fold branch point. In this case a radial geodesic,
     whose projection to the ($r,v$) surface hits the point $r=2k$, passes
     through it and returns to greater radii $r$ but with another value
     of $v$, thus leaving the particular 4-dimensional section of the
     $D$-dimensional space-time. However, if the multidimensional quantum
     wave function of the corresponding particle is $v$-independent, the
     particle does not disappear from an observer's sight and can look as if
     reflected from a mirror. The same true for macroscopic bodies if
     their energy-momentum is $v$-independent. If, on the contrary, the
     T-hole appears in a braneworld-like model, such that matter is
     concentrated at a particular value of $v$, then, being reflected from a
     T-horizon, matter disappears from the observers' sight.

     A T-hole is an example of a configuration looking
     drastically different in different conformal frames. If in \eq
     (\ref{g4}) the function $f$ is a multiple of $\sigma_2$ (it is natural
     since $\e^{\sigma_2}$ is the volume factor of extra dimensions; an
     example is (\ref{g*})), then the 4-metric (\ref{g4}) has a
     curvature singularity on the T-horizon. Indeed, the factor $\e^{2f}$
     is then proportional to a certain power of $g_{vv}$ which vanishes
     there. A consistent description of $\HT$ requires, however, the
     full multidimensional picture, where a curvature singularity is absent.

\section {Concluding remarks}

     We have obtained expressions for the Eddington PN parameters $\beta$
     and $\gamma$ for a wide range of \ssph\ solutions of multidimensional
     gravity with the general string-inspired action (\ref{2.1}).
     The existing experimental limits (\ref{obs-be}) and
     (\ref{obs-ga}) on $\beta$ and $\gamma$ constrain certain combinations
     of the solution parameters. This, however, concerns only
     the particular system for which the measurements are carried out, in
     our case, the Sun's gravitational field. The main feature of the
     expressions for $\beta$ and $\gamma$ is their dependence not only on
     the theory (the input constants entering into the action), but on the
     particular solution (the integration constants). This means that the PN
     parameters should be different for different self-gravitating
     configurations. They should not only be different, say, for stars and
     \bhs, but even for different stars if we try to describe their external
     fields in terms of the model (\ref{2.1}).

     A feature of interest is the universal prediction of $\beta > 1$ in
     (\ref{be*}) for both frames (\ref{4-E}) and (\ref{g*}). This
     conclusion does not depend on the system integrability and rests solely
     on the positivity of energy required.

     The predicted deviations of $\gamma$ from unity may be of any sign and
     depend on many integration constants. It turns out that
     precisely $\gamma=1$ in the external field of a BH with equal
     electric and magnetic charges of the same composite $F$-form or of two
     forms of equal ranks, or with a few such pairs of charges.

     If, however, the 4-dimensional Einstein frame is adopted as the
     observational one, we have a universal result $\gamma=1$ for all \ssph\
     solutions of the theory (\ref{2.1}).

     The BH temperature $\TH$ also carries information about the
     multidimensional structure of space-time. Being a universal parameter
     of a given solution to the field equations, $\TH$ as a function of the
     observable BH mass and charges is still conformal frame dependent due
     to different expressions for the mass $M$ in different frames.

     One more evident consequence of multidimensional theory is the Coulomb
     law violation, caused by a modification of the conventional Gauss
     theorem and also by scalar-electromagnetic interaction. A remarkable
     property of the modified Coulomb law is its conformal frame
     independence for any given \ssph\ metric where the electromagnetic
     field is situated.

     In addition to modifications of conventional physical laws, extra
     dimensions can lead to the existence of a new kind of objects,
     T-holes, which, as we argue, can probably be observable as bodies
     with mirror surfaces, at least if the T-horizons are connected with
     compact spacelike extra dimensions. It is also possible that matter
     simply escapes from our physical space across the T-hole surface. More
     detailed predictions can be formulated in specific theories.
     Although it seems hard to point out a T-hole formation mechanism which
     might act in the present Universe, their emergence should have been
     as probable as that of \bhs\ in the early Universe, when all
     space-time dimensions were on equal footing.

\subsection*{Acknowledgements}

This work was supported in part by the Russian Foundaation for Basic
Research and Project SEE. KB acknowledges partial financial support from
a NATO Science Fellowship Programme grant and kind hospitality at the
Dept.  of Mathematics of University of the Aegean, Karlovassi, Samos,
Greece, where part of the work was done. VNM is grateful to CONACYT, Mexico,
for partial financial support and to Depto. de Fisica, CINVESTAV, for
hospitality during his stay there.  We also thank Spiros Cotsakis, Vladimir
Ivashchuk and Dmitri Gal'tsov for helpful discussions.

\def\Jl#1#2{{\it #1\/} {\bf #2},\ }

\def\CQG#1 {\Jl{Clas. Qu. Grav.}{#1}}
\def\DAN#1 {\Jl{Dokl. AN SSSR}{#1}}
\def\GC#1 {\Jl{Grav. \& Cosmol.}{#1}}
\def\GRG#1 {\Jl{Gen. Rel. Grav.}{#1}}
\def\JETF#1 {\Jl{Zh. Eksp. Teor. Fiz.}{#1}}
\def\JMP#1 {\Jl{J. Math. Phys.}{#1}}
\def\NP#1 {\Jl{Nucl. Phys.}{#1}}
\def\PLA#1 {\Jl{Phys. Lett.}{#1A}}
\def\PLB#1 {\Jl{Phys. Lett.}{#1B}}
\def\PRD#1 {\Jl{Phys. Rev.}{D\ #1}}
\def\PRL#1 {\Jl{Phys. Rev. Lett.}{#1}}


\small
\begin{thebibliography}{99}

\bibitem{will}
	C.M. Will, ``Theory and Experiment in Gravitational Physics'',
	Cambridge University Press, Cambridge, 1993.

\bibitem{mel3}
	K.P. Staniukovich and V.N. Melnikov, ``Hydrodynamics, Fields and
	Constants in the Theory of Gravitation'',
	Energoatomizdat, Moscow, 1983, 256 pp. (in Russian).

\bibitem{we/nc88}
        K.A. Bronnikov, V.D. Ivashchuk and V.N. Melnikov,
% Time variation of the gravitational constant in multidimensional cosmology.
	{\it Nuovo Cim. } B102, 209 (1988).

\bibitem{mel4}
	V.N. Melnikov, {\it Int. J. Theor. Phys.} {\bf 33}, 1569 (1994).

\bibitem{mel5}
	V. de Sabbata, V.N. Melnikov and P.I. Pronin,
	{\it Prog. Theor. Phys.} {\bf 88}, 623 (1992).

\bibitem{mel6}
	V.N. Melnikov. In: ``Gravitational Measurements, Fundamental
	Metrology and Constants'', eds. V. de Sabbata and V.N. Melnikov,
	Kluwer Academic Publ., Dordtrecht, 1988, p. 283.

\bibitem{damr}
	T. Damour, ``Gravitation, experiment and cosmology", gr-qc/9606079;\\
	C. Will, ``The confrontation between general relativity
		and experiment", gr-qc/0103036.

\bibitem{alpha-}
	J.K. Webb, M.T. Murphy, V.V. Flambaum, V.A. Dzuba, J.D. Barrow, C.W.
	Churchill, J.X. Prochaska  and A.M. Wolfe,
	``Further evidence for cosmological evolution of the fine structure
	constant'', astro-ph/0012539.

\bibitem{zhuk}
	U. G\"unther, S. Kriskiv and A. Zhuk,
	\GC 4 1 (1998);\\
	U. G\"unther and A. Zhuk, \PRD {56} 6391 (1997);
	``Gravitational excitons as dark matter'', astro-ph/0011017.

\bibitem{GIM}
	V.R. Gavrilov, V.D. Ivashchuk and V.N. Melnikov,
	\JMP {36} 5829 (1995).

\bibitem{M1}
        V.N. Melnikov,
	``Multidimensional Cosmology and  Gravitation'',
	CBPF-MO-002/95, Rio de Janeiro, 1995, 210 pp.

\bibitem{bm-ann}
        K.A. Bronnikov and V.N.Melnikov,
%      ``Black holes in multidimensional dilaton gravity: existence and
%        stability.''
        {\it Ann. Phys. (N.Y.)} {\bf 239}, 40 (1995).

\bibitem{akama}
	K. Akama, ``Pregeometry'',
	{\it Lect. Notes Phys.\/} {\bf 176}, 267 (1982).

\bibitem{rubshap}
        V.A. Rubakov and M.E. Shaposhnikov,
%	``Do we live in a domain wall?'',
        {\it Phys. Lett.} {\bf 125B}, 136 (1983).

\bibitem{ransum}
	L. Randall and R. Sundrum,
%	``An alternative to compactification'',
	\PRL {83}, 4690 (1999); hep-th/9906064.

\bibitem{maia}
	M.D. Maia and V. Silveira, \PRD {48} 954 (1993).

\bibitem{IM0}
	V.D. Ivashchuk and V.N. Melnikov,
	{\it Grav. and Cosmol.\/} {\bf 2}, No 4 (8), 297 (1996);
	hep-th/9612089.

\bibitem{IM2}
	V.D. Ivashchuk and V.N. Melnikov,
	{\it Phys. Lett.\/} {\bf B 384}, 58 (1996).

\bibitem{IM5}
	V.D. Ivashchuk and V.N. Melnikov,
%%      ``Sigma-model for generalized  composite p-branes",
  	hep-th/9705036, {\it Class. and Quant. Grav.} {\bf 14}, 3001 (1997).

\bibitem{bim97}
  	K.A. Bronnikov, V.D. Ivashchuk and V.N. Melnikov,
	{\it Grav. and Cosmol.\/} {\bf 3}, 203 (1997);
	gr-qc/9710054.

\bibitem{bkr97}
        K.A. Bronnikov, U. Kasper and M. Rainer,
%       ``Spherically symmetric solutions for intersecting electric and
%       magnetic $p$-branes'',
        \GRG {31} 1681 (1999);
        gr-qc/9708058.

\bibitem{GSW}
	M.B. Green, J.H. Schwarz, and E. Witten, ``Superstring
	Theory" in 2 vols. (Cambridge Univ. Press, 1987).

\bibitem{brane}
C. Hull and P. Townsend,
%	``Unity of superstring dualities",
	{\it Nucl. Phys.\/} {\bf B 438}, 109 (1995);\\
P. Horava and E. Witten, {\it Nucl. Phys.\/} {\bf B 460}, 506 (1996),
	hep-th/9510209; hep-th/9603142;\\
J.M. Schwarz,  ``Lectures on superstring and M-theory dualities",
	{\it Nucl. Phys. Proc. Suppl.\/} {\bf 55B}, 1 (1997);
	hep-th/9607201;\\
K.S. Stelle, ``Lectures on supergravity \branes'', hep-th/9701088;\\
M.J. Duff,
%	``M-theory (the theory formerly known as strings)'',
	{\it Int. J. Mod. Phys.\/} {\bf A 11}, 5623 (1996);
	hep-th/9608117.

\bibitem{im2t}
    V.D. Ivashchuk  and  V.N. Melnikov,
    {\it Class. and Quant. Grav.} {\bf 11}, 1793 (1994);
    {\it Int. J. Mod. Phys. D} {\bf 4}, 167 (1995).

\bibitem{2times}
	I. Bars and C. Kounnas,
% 		``String and particle with two times",
		\PRD {56} 3664 (1997);
		hep-th/9703060;    \\
	H. Nishino,
%	       ``Supergravity in 10+2 dimensions
% 		as a consistent background for superstring",
		\PLB {428} 85 (1998);
		hep-th/9703214.

\bibitem{khuri}
	C.M. Hull and R.R. Khuri,
%	``Worldvolume theories, holography, duality and time'',
	\NP {B 575} 231 (2000)
	hep-th/9911082.

\bibitem{khv}
	N. Khviengia, Z. Khviengia, H. L\"u and C.N. Pope,
%	``Toward field theory of F-theory'',
	\CQG {15} 759 (1998);
	hep-th/9703012.

\bibitem{im-cots}
	S. Cotsakis, V.D. Ivashchuk and V.N. Melnikov,
	\GC 5 52 (1999); gr-qc/9902148.

\bibitem{im-mank}
	V.D. Ivashchuk, V.S. Manko and V.N. Melnikov,
	\GC 6 219 (2000).

\bibitem{br95-1}
    	K.A. Bronnikov,
%    	``Extra dimensions and possible space-time signature changes'',
    	{\it Int. J. Mod. Phys. D\/}, {\bf 4}, 4, 491 (1995).

\bibitem{br95-2}
    	K.A. Bronnikov,
%    	``Spherically symmetric solutions in D-dimensional dilaton gravity'',
    	{\it Grav. \& Cosmol.\/} {\bf 1}, 1, 67 (1995).

\bibitem{bobs}
	K.A. Bronnikov,
%	``Block-orthogonal brane systems, black holes and wormholes",
	{\it Grav. and Cosmol.\/} {\bf 4}, 49 (1998);
	hep-th/9710207.

\bibitem{br-jmp}
	K.A. Bronnikov,
%	``Gravitating brane systems: some general theorems",
	{\it J. Math. Phys.\/} {\bf 40}, 924 (1999);
	gr-qc/9806102.

\bibitem{br-73}
	K.A. Bronnikov,
        {\it Acta Phys. Polon. } {\bf B4}, 251 (1973).

\bibitem{IMmapa}
        V.D. Ivashchuk and V.N. Melnikov,
%%	``Madjumdar-Papapetrou solutions in sigma model and intersecting
%% 	p-branes''
	\CQG {16} 849 (1999); hep-th/9802121.

\bibitem{im9901}
	V.D. Ivashchuk and V.N. Melnikov,
	{\it in:\/} Proc. 2nd Samos Meeting, 1998;
	{\it Lecture Notes in Physics,\/} v. 537,
	``Mathematical and Quantum Aspects of Relativity and Cosmology'',
	eds. S. Cotsakis and G. Gibbons, Springer, Berlin, 2000, p. 214;
	gr-qc/9901001.

\bibitem{im9910}
	V.D. Ivashchuk and V.N. Melnikov,
%	``Black-hole $p$-brane solutions for general intersection rules'',
	\GC {6} 27 (2000); \CQG {17} 2073 (2000); hep-th/9910041.

\bibitem{horst91}
	G.T. Horowitz and A. Strominger,
	\NP{B 360} 197 (1991).

\bibitem{bm-stab}
	K.A. Bronnikov and V.N. Melnikov,
%	``p-Brane black holes as stability islands'',
	\NP{B 584} 436 (2000).

\bibitem{glaf94}
	R. Gregory and R. Laflamme,
	\NP{B 428}, 399 (1994).

\bibitem{sokol93}
	G. Magnano and L.M. Sokolowski,
	\PRD{50} 5039 (1994); gr-qc/9312008.

\bibitem{fara98}
	V. Faraoni, E. Gunzig and P. Nardone,
	``Conformal transformations in classical gravitational theories
	and in cosmology'', gr-qc/9811047;
	{\it Fundamentals of Cosmic Physics } {\bf 20}, 121 (1999).

\bibitem{rzhuk98}
	M. Rainer and A.I. Zhuk,
	\GRG{32} 79 (2000); gr-qc/9808073.

\bibitem{noji01}
	S. Nojiri, O. Obregon, S.D. Odintsov and V.I. Tkach,
	``String versus Einstein frame in  AdS/CFT induced
	quantum dilatonic brane-world Universe'',
	hep-th/0101003.

\bibitem{banks}
	T. Banks and M. O'Loughlin, \PRD {47} 540 (1993).

\bibitem{shira}
        K. Shiraishi, {\it Mod. Phys.  Lett. A} {\bf 7} (1992),
        3449; 3569; {\it Phys. Lett.} {\bf 166A}, 298 (1992).

\bibitem{Reas}
%%      R.D. Reasenberg et. al., {\it Astrophys. J. } {\bf 234}, L219 (1979).
	T.M. Eubanks at al.,
	``Advances in solar system tests of gravity'',
	preprint, available at\\
	ftp://casa.usno.navy.mil/navnet/postscript/, file prd\_15.ps (1999).

\bibitem{Dik}
        J.O. Dickey et al., {\it Science } {\bf 265}, 482 (1994).

\bibitem{Nor}
        K. Nordtvedt, {\it Phys. Rev. } {\bf 169}, 1017 (1968).

\bibitem{wald}
	R. Wald, ``General Relativity'', Univ. of Chicago Press, Chicago,
	1984.

\bibitem{JaK}
	T. Jacobson and G. Kang, \CQG {10} L201 (1993).

\bibitem{groper}
        D.J. Gross and M. Perry, {\it Nucl. Phys.} {\bf B226}, 29 (1983).

\bibitem{br79}
        K.A. Bronnikov,
	{\it Izvestiya Vuzov, Fizika}, 1979, No.\,6, 32 (in Russian).

\end{thebibliography}
\end{document}